\UseRawInputEncoding

\documentclass[a4paper,fleqn]{cas-sc}
\usepackage{makecell}
\usepackage{algorithm}  
\usepackage{algorithmic}
\usepackage[mathlines, pagewise]{lineno}

\usepackage[numbers, compress]{natbib}

\def\tsc#1{\csdef{#1}{\textsc{\lowercase{#1}}\xspace}}
\tsc{WGM}
\tsc{QE}

\begin{document}
\let\WriteBookmarks\relax
\def\floatpagepagefraction{1}
\def\textpagefraction{.001}

\shorttitle{A Wavelet Transform and self-supervised learning-based framework for bearing fault diagnosis with limited labeled data}

\shortauthors{Y. Jin, L. Hou, M. Du and Y. Chen}

\title [mode = title]{A Wavelet Transform and self-supervised learning-based framework for bearing fault diagnosis with limited labeled data}

\tnotemark[1]

\tnotetext[1]{Supported by the National Natural Science Foundation of China (No. 11972129) and the National Major Science and Technology Projects of China (No. 2017-IV-0008-0045).}

%

\author[]{Yuhong Jin}[type=editor,
      style=chinese
]







\author[]{Lei Hou}[type=editor,
      style=chinese,
      orcid=0000-0003-0271-7323
]

\cormark[1]


\ead{houlei@hit.edu.cn}

\ead[url]{http://homepage.hit.edu.cn/houlei}

\author[]{Ming Du}[type=editor,
      style=chinese
]

\author[]{Yushu Chen}[type=editor,
      style=chinese
]



\cortext[1]{Corresponding author}


\affiliation[]{organization={School of Astronautics},
      addressline={Harbin Institute of Technology},
      city={Harbin},
      postcode={150001},
      country={P. R. China}}

\begin{abstract}
      Traditional supervised bearing fault diagnosis methods rely on massive labeled data, yet annotations may be very time-consuming or infeasible. The fault diagnosis approach that utilizes limited labeled data is becoming increasingly popular. In this paper, a Wavelet Transform (WT) and self-supervised learning-based bearing fault diagnosis framework is proposed to address the lack of supervised samples issue. Adopting the WT and cubic spline interpolation technique, original measured vibration signals are converted to the time-frequency maps (TFMs) with a fixed scale as inputs. The Vision Transformer (ViT) is employed as the encoder for feature extraction, and the self-distillation with no labels (DINO) algorithm is introduced in the proposed framework for self-supervised learning with limited labeled data and sufficient unlabeled data. Two rolling bearing fault datasets are used for validations. In the case of both datasets only containing 1\% labeled samples, utilizing the feature vectors extracted by the trained encoder without fine-tuning, over 90\% average diagnosis accuracy can be obtained based on the simple K-Nearest Neighbor (KNN) classifier. Furthermore, the superiority of the proposed method is demonstrated in comparison with other self-supervised fault diagnosis methods. 
\end{abstract}


\begin{highlights}
      \item A novel fault diagnosis framework for bearings with limited labeled data is developed. 
      \item Satisfactory diagnosis accuracy is achieved on two bearing fault datasets consisting of only 1\% labeled data. 
      \item The effect of hyperparameters on fault diagnosis performance and computational complexity of the proposed framework is discussed in detail.
      \item Without any supervised sample, fault-specific features with inter-class separability are observed via visualization technique. 
\end{highlights}


\maketitle

\section{Introduction}

Rotating machinery plays a vital role in modern industries, so its condition monitoring and health management are of great importance. According to statistics, about 45\%-55\% of rotating machinery and equipment failure is caused by damage to the bearing part \cite{Nandi}. Therefore, timely and accurate bearing fault diagnosis has always been highly demanded to enhance machine reliability \cite{ZhaoXL, LIU2020106518}. Traditional bearing fault diagnosis methods are based on the physical model, and the specific fault components are analyzed by various signal processing techniques \cite{JIANG201836, GLOWACZ2021108070, ZhijianWang222}. However, these physical models are not universal in complex, high-dimensional systems. With the rapid development of industrial technology, the structure of rotating machinery becomes more and more complicated, and the limitations of conventional fault diagnosis methods are gradually highlighted. 

In recent years, benefited from the advances in industrial computer and sensor technology, data-driven intelligent fault diagnosis method has attracted more and more attention from researchers due to the great merits of high accuracy and low requirement for prior knowledge \cite{WANG2021591}. Data-driven fault diagnosis methods can be divided in two categories: Machine Learning (ML) based approach and Deep Learning (DL) based approach. Currently, many ML models such as Support Vector Machine (SVM) \cite{YAN201847, HAN2021109022, YUAN12548}, Self-Organized Map (SOM) \cite{Xiao, FAN56584}, Auto-Encoder (AE) \cite{MAO2021107233, YANG2021133}, and Radial Basis Function (RBF) neural network \cite{WOS:000536474500001, WOS:000488645700003} have been extensively employed in the field of fault diagnosis. Nevertheless, most of these approaches still require manual design features, which means limitations in adaptive feature extraction. 

DL based techniques can automatically learn the feature representation from mass data according to the given task and have a strong feature extraction ability. These approaches have been widely developed, and numerous promising results have been acquired \cite{DING2020}. Assorted neural network architectures and enhanced techniques, such as Deep Belief Network (DBN) \cite{YANG2021104149}, Convolutional Neural Network (CNN) \cite{KIRANYAZ2021107398, 9343326, WOS:000713937700001}, Recurrent Neural Network (RNN) \cite{LIU2021110214, 9056473}, Attention mechanism \cite{LI2022110500}, Transformer \cite{9579015, DU2022110545} and their variants \cite{Wang_Lei_Yan_Li_Nandi_2021, ZHAO2022} have also been generally exploited for the fault diagnosis. Jie et al. \cite{WOS:000749744000003} established a novel Gaussian-Bernoulli deep belief network (GDBN) model for intelligent fault diagnosis, where the graph regularization and sparse features learning are embedded. Combining the advantages of attention mechanism, Squeeze-and-Excitation Network (SENet) \cite{8701503}, and soft threshold, Zhao et al. \cite{8850096} developed the Deep Residual Shrinkage Networks (DRSN). This new deep learning model can achieve a high fault diagnosis accuracy under strong noise interference. Aiming at the problem that the connection of the local fragments (namely quasi-periodicity) in the measured vibration signals is easy to be neglected, Gao et al. \cite{GAO2021107413} performed a novel weak fauld diagnosis method for the rolling bearings based on the Long Short Term Memory (LSTM) network and multichannel Continuous Wavelet Transform (MCCWT). Compared to the traditional CWT, MCCWT converts the original sample space into a multichannel representation, which improves the feature extraction capability of the LSTM. In recent work, Ding et al. \cite{DING2022108616} applied the Transformer architecture to fault diagnosis of rolling bearings. Based on the view that the time-frequency map obtained by signal processing is formed by splicing the instantaneous spectrum of the signal over a period of time, the time-frequency Transformer (TFT) model is presented. Through experiments, the effectiveness and superiority of the proposed TFT are validated. In brief, the advances in DL based fault diagnosis methods fully prove the potency of the emerging data-driven algorithm for processing complex mechanical systems \cite{Hoang}. 

Satisfactory fault diagnosis results reported in the above literature are premised on the sufficient labeled data. These DL based approaches are based on the paradigm of supervised learning for end-to-end training, thus acquiring the features with generalization ability \cite{ZHAO2019213}. However, in real engineering environment, labeling large amounts of measured data may be time-consuming and costly. Besides, considering the complexity and uncertainty of the machinery system, in many cases, the fault mode corresponding to the acquired vibration signal is virtually unidentified, which leads that supervised learning is not effective enough. Therefore, the problem of fault diagnosis with limited labeled data has aroused extensive attention from researchers. Currently, two main approaches to this problem are dataset extension and exploring unlabeled data. Dataset extension aims to generate the additional "labeled" data based on the limited labeled samples through a technique similar to interpolation, including various data augmentation methods \cite{BAI2021109885, ZHANG201834}, Generative Adversarial Network (GAN) \cite{ZHENG2020107741, 9606743, Wang_Liu_Ai_2022}, etc. Li et al. \cite{LI458623} designed a data augmentation method that combined different signal processing techniques such as masking noise, signal translation, stretching, etc. The experimental results show that the diagnosis performance of the DL model can get a promotion from more generated samples. By employing the Sparse Auto-Encoder (SAE) to reduce the dimension of the original data, Ma et al. \cite{MA2021115234} improved the traditional GAN and proposed the Sparsity-Constrained Generative Adversarial Network (SCGAN), which better converges to Nash equilibrium, and higher diagnosis accuracy can be achieved with limited labeled data. Different from the above research, Wang et al. came up with a novel dataset extension method based on the Sub-Pixel Convolutional Neural Network (ESPCN) \cite{8825083}. Through this method, generated data with high-resolution can be acquired. Experimental results of gearbox and bearing datasets show that the proposed method has strong feasibility to carry out data augmentation for fault diagnosis of rotating machines under the speed fluctuation condition. Overall, sample size can be enlarged using dataset extension, and the lack of labeled data is alleviated. 

However, fake samples generated based on the dataset extension technology will inevitably be similar to the real samples, resulting in lacking data diversity, which will increase the risk of overfitting. In addition, mass unlabeled data without precise machine health conditions, which can be easily collected in general, does not get effective utilization in these methods. 
To solve these problems, fault diagnosis methods under the limited labeled samples case by exploring unlabeled data are presented, known as unsupervised learning \cite{9290051, LIU2021107488}, self-supervised learning \cite{DING2022108126, WOS:000751119800001, WOS:000563132100002}, and semi-supervised learning \cite{9270010, WOS:000582808200002, JIAN2021104365}. Zhang et al. \cite{ZHANG2021106679} offered an unsupervised framework with Reconstruction Sparse Filtering (RSF) for rolling bearing diagnosis. The basis vectors are constrained explicitly by a Soft-Reconstruction Penalty (SRP), enabling RSF to learn a group of independent basis vectors to extract dissimilar features without applying any labeled sample. Li et al. \cite{LI2022108663} leveraged deep InfoMax (DIM) to improve the generalization ability of the features extracted by a CNN encoder, which can alleviate the overfitting problem when the labeled fault samples are limited. Li et al. \cite{LI2020106825} designed a three-stage semi-supervised fault diagnosis  conjoin unsupervised clustering and supervised fine-tuning. Even only one labeled sample for each class, a high testing accuracy is obtained by this method. The above research results demonstrate the feasibility of exploring unlabeled data to address the challenging diagnostic tasks with limited labeled samples.

This paper focuses on the problem of bearing fault diagnosis with availabilities of limited labeled data and sufficient unlabeled data. Under this proposition, a Wavelet Transform and self-supervised learning-based framework is proposed. Herein, the Vision Transformer is adopted as an encoder to extract the feature vectors, which possesses the preponderance of great parallelism and strong scalability. Accordingly, the projector head is introduced on the top of the encoder to establish the model network for self-supervised learning, whose output is pseudo labels. Based on this, a pretext task called "local to global correspondence" is constructed by initializing the teacher and student networks with the same architecture. Moreover, the centering and sharpening operations included in the self-distillation with no labels algorithm are integrated into the proposed framework to train the model networks, which enable us effectively avoid the mode collapse during the training procedure. Furthermore, the Wavelet Transform technique is used to convert the original time-domain vibration signals into time-frequency maps. This preprocessing method helps the encoder learn a better feature representation. Experiments on two bearing fault datasets where only 1\% labeled data is contained are implemented to validate the proposed method, and over 90\% average testing accuracy is achieved. 

\section{Proposed method}

The proposed method mainly comprises three major parts: (1) preprocessing pipeline of the raw vibration signals; (2) model network for feature representation; (3) training algorithm for the model network. In this section, we will illustrate our proposed methodology in detail.

\subsection{Wavelet Transform and preprocessing pipeline}

\begin{figure}[h]
      \centering
      \includegraphics[width = 0.99\textwidth]{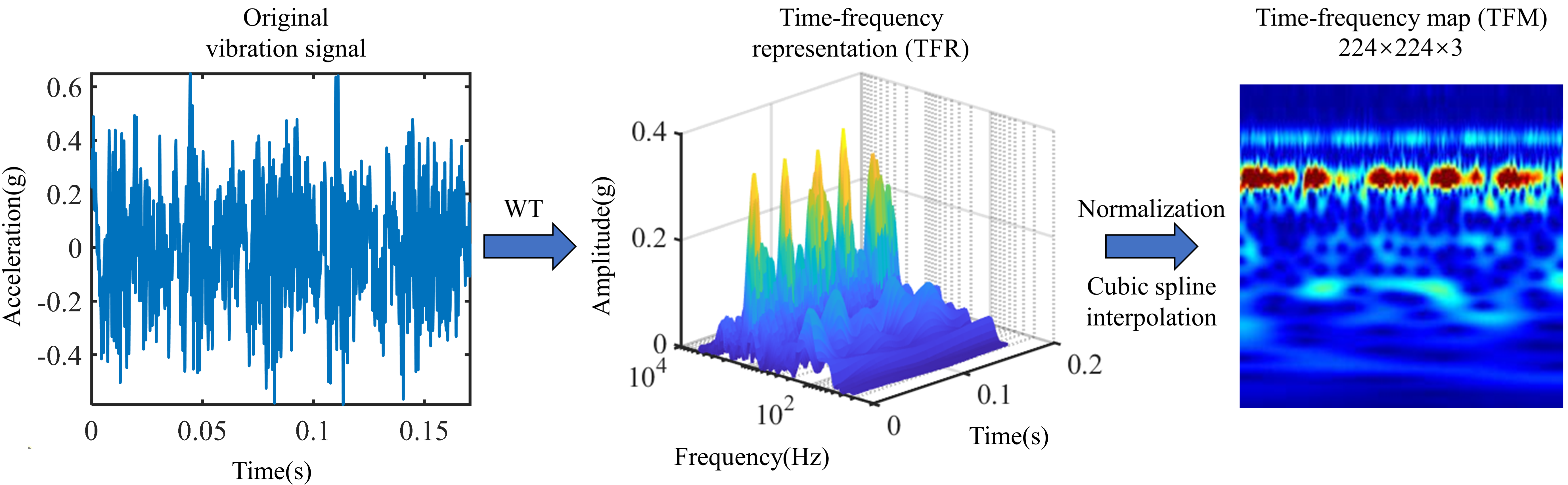}
      \caption{Preprocessing pipeline of the original vibration signal}
      \label{fig:preprocess}
\end{figure}

In the fault diagnosis of rotating machinery, background noise is pervasive due to sensors' noise input and environmental factors, which will interfere with the time-domain signals. Additionally, the acquired vibration signals are usually non-stationary due to speed fluctuation and fault. In this case, neither time-domain analysis nor frequency-domain analysis is suitable. Therefore, it is necessary to convert the raw vibration signals into time-frequency representation (TFR). In this paper, we utilize the Wavelet Transform (WT) as the time-frequency domain analysis method. The WT of a given vibration signal $x(t)$ can be defined as
\begin{equation}
      WT(a, \tau) = \frac{1}{\sqrt{a}}\int_{-\infty}^{+\infty}x(t)\psi(\frac{t-\tau}{a})dt = \int_{-\infty}^{+\infty}x(t)\psi_{a, \tau}(t)dt
      \label{eq:WT}
\end{equation}
where $a$ is scaling parameter, $\tau$ is time translation parameter. $\psi_{a, \tau}(t) = \frac{1}{\sqrt{a}}\psi(\frac{t-\tau}{a})$ is wavelet basis function of the WT. Fig.\ref{fig:preprocess} shows the TFR of the raw vibration signals based on the WT. We can find that the frequency-domain signals of each time step are well extracted. Nevertheless, it should be noted that the size of TFR is related to the number of sampling points, which means that the shape of TFR may be uncertain. Besides, numerous sampling points can result in a huge TFR, making subsequent calculations difficult. To fix the shape of TFR, as shown in Fig.\ref{fig:preprocess}, we introduce a resize method. The specific approach is as follows: first, the amplitude range of TFR is scaled to [0, 1] by normalization. Then, the amplitude values of TFR are mapped to color values. Finally, the size of TFR is fixed at $224 \times 224$ by cubic spline interpolation. The TFR after the above treatments is denoted as time-frequency map (TFM), whose shape is $224 \times 224 \times 3$. TFM can be directly used as input to the model network.

\subsection{Model network}

\subsubsection{Encoder}

\begin{figure}[h]
      \centering
      \includegraphics[width = 0.99\textwidth]{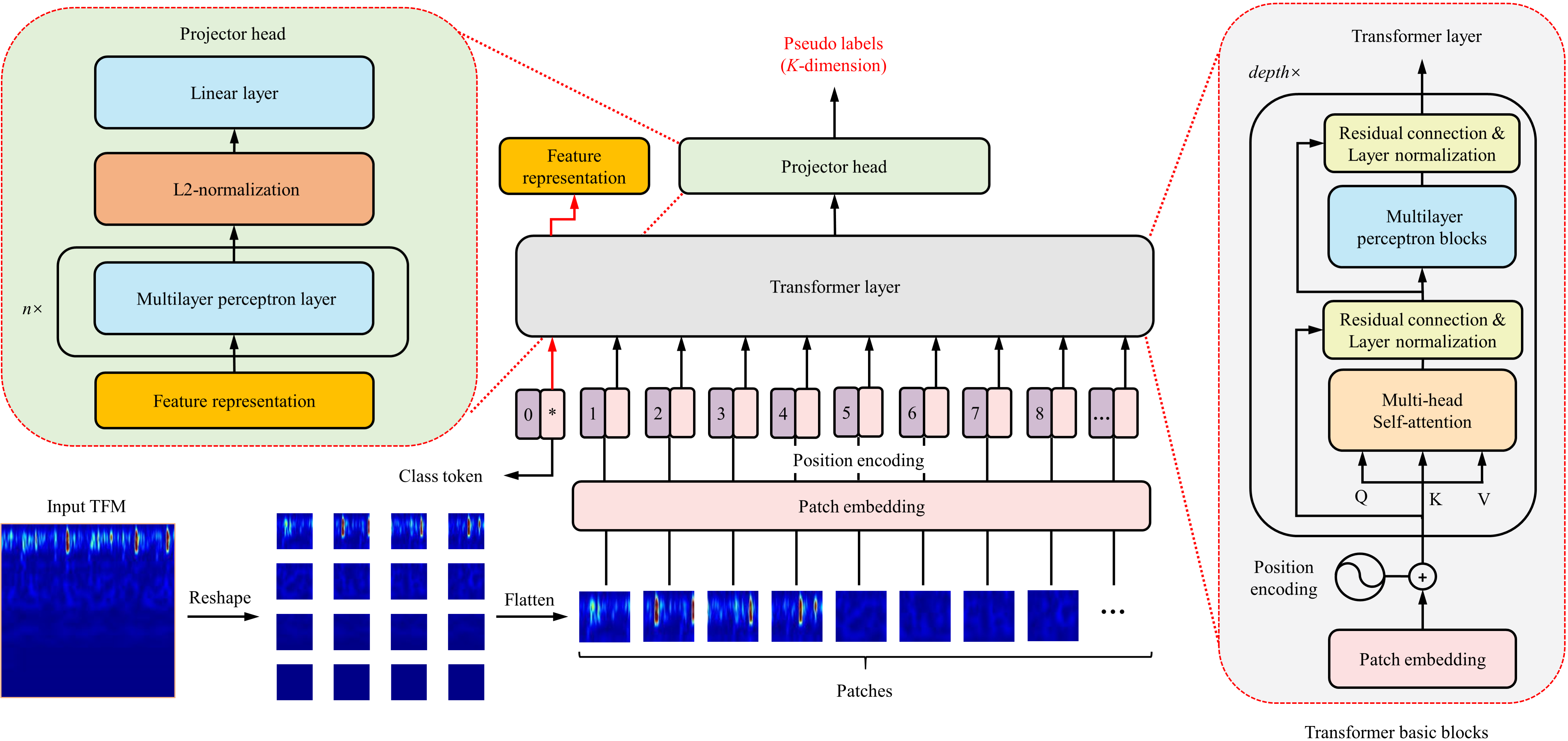}
      \caption{The architecture of the model network.}
      \label{fig:model_network}
\end{figure}

We need two networks in the following training process: teacher network and student network. In this paper, both of them are collectively called the model network. As shown in Fig.\ref{fig:model_network}, the model network is consists of two stages: an encoder and a projector head. The encoder provides a feature extraction and representation from the input TFM. A variety of DL models, such as Residual Neural Network (ResNet) and Vision Transformer (ViT) \cite{ViT}, can be used as encoder's backbone, and in this paper, we employ the ViT as our backbone. The impact of different backbones on the fault diagnosis performance will be discussed later. For the sake of presentation, the input TFM is denoted as $x \in \mathbb{R}^{H \times W \times C}$ ignoring the batch size, where $(H, W)$ is the shape of the TFM, $C$ is the number of channels. As described previously, $H = W = 224$, and $C$ is 3. To conform the input form of the standard Transformer, we reshape the TFM into a sequence of flattened 2D patches, denoted as $x_{p} \in \mathbb{R}^{{N}\times ({P^2 \times C})}$， where $(P, P)$ is the resolution of each TFM patch, $N$ is the number of patches, and $N = HW/P^2$. This process is visually illustrated in Fig.\ref{fig:model_network}, which divides the input TFM evenly into patches of a specified size. Then, we reorder these patches into a sequence, as shown in Eq.\eqref{eq:reshape}
\begin{equation}
      x \in \mathbb{R}^{H \times W \times C} \stackrel{\text{reshape}}{\longrightarrow} x_{p} = [x_{p}^{1}, x_{p}^{2}, ..., x_{p}^{N}] \in \mathbb{R}^{{N}\times ({P^2 \times C})}
      \label{eq:reshape}
\end{equation}
where $x_{p}^{i}$ is the $i$th patch, and $i = 1, 2, 3, ..., N$. Then, like the procedure of word embedding, we flatten each patch and map them to the high-dimensional embedding space through a learnable linear projection, as given in Eq.\eqref{eq:embedding}
\begin{equation}
      z_{0} = [x_{p}^{1}, x_{p}^{2}, ..., x_{p}^{N}] \cdot W_{emd} \in \mathbb{R}^{N \times d}
      \label{eq:embedding}
\end{equation}
where $W_{emd} \in \mathbb{R}^{(P^{2} \times C ) \times d}$ is a trainable embedding matrix, $d$ is the embedding dimension. It should be noted that all patches share the same embedding matrix, so Eq.\eqref{eq:embedding} is mathematically equivalent to a 2D convolution operation. Then, since the Transformer does not contain the position information, position encoding is added to retain the absolute and relative position information of the patches. And beyond that, similar to BERT's class token, we present a learnable vector (denoted as $x_{class}$) to serve as the feature representation, which can be obtained by Eq.\eqref{eq:position_encoding_class}
\begin{equation}
      z_{0} \leftarrow [x_{class}; z_{0}] + E_{pos}
      \label{eq:position_encoding_class}
\end{equation}
where $x_{class} \in \mathbb{R}^{d}$, $E_{pos} \in \mathbb{R}^{(N+1) \times d}$. Both of them are learnable parameters. Eq.\eqref{eq:reshape} to Eq.\eqref{eq:position_encoding_class} are collectively called patch embedding, as shown in Fig.\ref{fig:model_network}. $z_{0}$ is referred to as "embedding sequence."

The following part of the encoder can be regarded as a feature extraction structure, which takes the embedding sequence $z_{0}$ as input. In the ViT architecture, the core part of feature extraction is the Transformer layer, which comprises multiple Transformer basic blocks stacked on top of each other. Fig.\ref{fig:model_network} introduces the structure of the Transformer base blocks, including Multi-head Self-attention mechanism (MSA) and Multilayer perceptron blocks (MLP). MSA is the most  critical definition in Transformer basic blocks, which is based on the attention mechanism. The attention mechanism can be explained in terms of soft addressing. Consider that an element in the memory is composed of a key ($K$) and value ($V$). Currently, there is a query ($Q$), and we need to pick out $V$ based on the similarity between $Q$ and $K$. Note that for the same $Q$ value, instead of hard addressing, we may take all the $V$s in the memory and sum them weighted based on their importance. The importance of $V$ is measured by the comparability between $Q$ and its corresponding $K$, which is denoted as the Attention Distribution (AD). The calculation method of AD is called the score function, such as Scaled Dot-Product Attention, Bahdanau Attention, and Content-based Attention. In this paper, we adopt the Scaled Dot-Product Attention in our backbone, which is described as
\begin{equation}
      attn(Q, K, V) = softmax(\frac{QK^{T}}{\sqrt{d_{K}}})V
      \label{eq:attn}
\end{equation}
where $d_{K}$ is the embedding dimension of $K$. $\frac{1}{d_{K}}$ is a scaling factor for stabilizing the gradient. However, the attention mechanism in Eq.\eqref{eq:attn} cannot extract the information in the embedding sequence under the different subspace. Then, to avoid this deficiency, similar to the group convolution operation, researchers have introduced the Multi-head attention (MHA)mechanism, which can be given as
\begin{equation}
      \begin{aligned}
            MHA(Q, K, V)             & = [head_{1}, head_{2},..., head_{h}] \cdot W_{O} \\
            \text{where} \; head_{i} & = attn(QW_{Q}^{i}, KW_{K}^{i}, VW_{V}^{i})
      \end{aligned}
      \label{eq:multi_head_attn}
\end{equation}
where $W_{O} \in \mathbb{R}^{hd_{V} \times d}$, $W_{Q}^{i} \in \mathbb{R}^{d \times d_{K}}$, $W_{K}^{i} \in \mathbb{R}^{d \times d_{K}}$, $W_{V}^{i} \in \mathbb{R}^{d \times d_{V}}$, $h$ is the number of $head$ and $d_{V}$ is the dimension of values. Furthermore, when $Q = K = V$, Eq.\eqref{eq:multi_head_attn} is also called Multi-head Self-attention mechanism, namely MSA. The output of MSA in $l$th Transformer basic block (denoted as $z_{l}^{MSA}$) with residual connection and Layer normalization (LN) can be described as
\begin{equation}
      \begin{aligned}
            z_{l}^{MSA} & = MSA(LN(z_{l-1})) + z_{l-1}                           \\
                        & = MHA(LN(z_{l-1}), LN(z_{l-1}), LN(z_{l-1})) + z_{l-1}
      \end{aligned}
      \label{eq:MSA}
\end{equation}
where $z_{l}$, $l = 1, 2, 3, ..., depth$ is the output of the $l$th Transformer basic block in Transformer layer, and $depth$ is the number of Transformer blocks. $z_{0}$ is the embedding sequence.

Moreover, MLP blocks are applied after MSA in every Transformer basic block to achieve more complex nonlinear mapping. The MLP block in the $l$th Transformer basic block includes a nonlinear projection layer with an activation function and a linear projection layer, which is given in Eq.\eqref{eq:MLP}
\begin{equation}
      MLP(z_{l}^{MSA}) = GeLU(z_{l}^{MSA}W_{1}^{l}+b_{1}^{l})W_{2}^{l}+b_{2}^{l}
      \label{eq:MLP}
\end{equation}
where $W_{1}^{l} \in \mathbb{R}^{d \times d_{MLP}}$, $b_{1}^{l} \in \mathbb{R}^{d_{MLP}}$, $W_{2}^{l} \in \mathbb{R}^{d_{MLP} \times d}$, $b_{2}^{l} \in \mathbb{R}^{d}$. $d_{MLP}$ represents the embedding dimension of the nonlinear projection layer, and $GeLU(x)$ indicates the Gaussian error Linear Unit (GeLU), which can be shown as Eq.\eqref{eq:GeLU}
\begin{equation}
      GeLU(x) = x \phi (x) = x[1 + erf(x/{\sqrt{2}})]/2 \approx 0.5x(1 + tanh[\sqrt{2/\pi}(x + 0.045x^3)])
      \label{eq:GeLU}
\end{equation}
where $\phi (x)$ is the standard Gaussian distribution function. GeLU is smoother over the entire input range than Rectified Linear Unit (ReLU) and Leaky ReLU, with no discontinuous gradient at 0. Combining the residual connection and LN, the output of MLP in the $l$th Transformer basic block (namely $z_l^{MLP}$) can be obtained as
\begin{equation}
      z_{l} = MLP(LN(z_{l}^{MSA})) + z_{l}^{MSA}
      \label{eq:MLP_block}
\end{equation}

Based on Eq.\eqref{eq:MSA} and Eq.\eqref{eq:MLP_block}, the final output of the Transformer layer and the feature representation (denoted as $y$) in the ViT are shown in
\begin{equation}
      \begin{aligned}
            z_{l}^{MSA} & = MSA(LN(z_{l-1})) + z_{l-1},         & l = 1, 2, 3,..., depth \\
            z_{l}       & = MLP(LN(z_{l}^{MSA})) + z_{l}^{MSA}, & l = 1, 2, 3,..., depth \\
            y           & = LN(z_{depth}[0])
      \end{aligned}
      \label{eq:TransformerLayerOut}
\end{equation}
where $z_{depth}$ is the final output of the Transformer layer, and $z_{depth}^{0}$ means the class token in $z_{depth}$. To facilitate the expression in the following, we denote the set of learnable parameters in the model network as $\theta$ and denote the calculation process of the encoder part as $f_{\theta}$. Then the feature representation can be expressed as $y \triangleq f_{\theta}(x)$, where $x$ is the input TFM.

\subsubsection{Projector head}

The architecture of the projection head is straightforward, consisting of an $n$-layer MLP, an L2-normalization layer, and a weight normalized linear layer. The calculation of the $n$-layer MLP and linear layer is almost precisely the same as Eq.\eqref{eq:MLP}, not tired in words here. The L2-Normalization layer stabilizes the training process. Overall, the projector head plays almost the same role as the last layers of the various supervised learning backbones, which transforms the feature representation of the encoder into the probability distribution over $K$ dimension. The difference is that this probability distribution is meaningless, which means that the projector head outputs the pseudo labels. The hidden and bottleneck dimension of the MLP in the projector head is expressed as $d_{MLP}^{head}$. Finally, similar to the expressive method in the encoder part, the calculation process of the pseudo labels is denoted as $g_{\theta}(y) = g_{\theta}(f_{\theta}(x)) = q_{\theta}(x)$, where $q = g \circ f$, and $y$ is the feature representation.

\subsection{Self-distillation with no labels}

\begin{figure}[h]
      \centering
      \includegraphics[width = 0.62\textwidth]{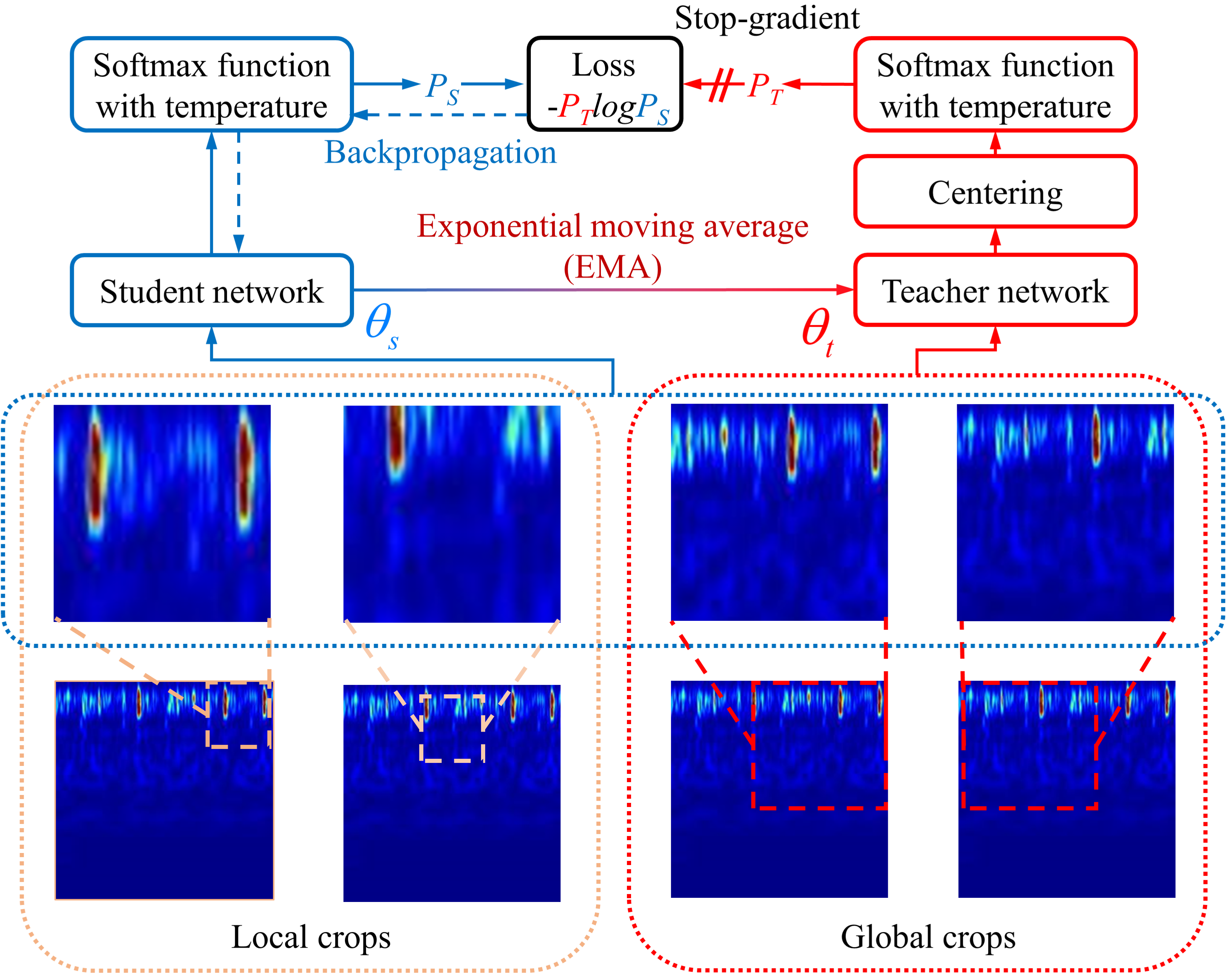}
      \caption{DINO's overall process.}
      \label{fig:DINO}
\end{figure}

When the sample data has no labels or only a few labels, traditional supervised learning cannot effectively train the network. We adopt the self-distillation with no labels (DINO) algorithm \cite{9709990} to train the model network for the fault diagnosis in the case of few labels. DINO is a self-supervised learning framework that can be interpreted as a form of knowledge distillation. DINO's goal is to make the network learn a feature representation that can be used for downstream tasks by exploring the information of the unlabeled data. As described previously, DINO uses two model networks to learn: the teacher and student networks, as shown in Fig.\ref{fig:DINO}. The teacher and student networks share the same architecture (shown in Fig.\ref{fig:model_network}), but parameterized by $\theta_{s}$ and $\theta_{t}$ respectively. Then, we design a self-supervised learning task called "local to global correspondence" to train the teacher and student networks. As shown in Fig.\ref{fig:DINO}, given a TFM, we can construct different views, or called crops, of the original image through the data augmentation (including random crop, Gaussian blur, color jittering, and solarization). Then, it should be noted that two different random crop scale parameters are utilized to obtain the global and local views separately. Each local crop contains only a small area (scale range [0.05, $s$]), while the global crops cover a large area (scale range [$s$, 1]) of the original TFM. $s$ is the scale parameter in the random crop process. In a random crop process, we can obtain a set of different crops, namely $X'$, which contains 2 global views (denoted as $X^{g} = \left\{x_{1}^{g}, x_{2}^{g}\right\}$) and $N$ local views (denoted as $X^{l} = \left\{ x_{1}^{l}, x_{2}^{l}, ..., x_{N}^{l} \right\}$), $X' = X^{l}\cup X^{g}$. The teacher network only accepts global crops ($X^{g}$) as input, while the student network passes all crops ($X'$).

Next, we will discuss the other computational details in DINO's algorithm. We want the student network's output to match the given teacher network. From this perspective, DINO is very similar to knowledge distillation \cite{562254848}. The difference lies in the teacher network in knowledge distillation is an extensive pre-trained network. In DINO, the teacher and student networks share the same structure and have not undergone any pre-training. Then, the probability of pseudo labels $P$ can be obtained by normalizing the output of the model network through a softmax function with temperature
\begin{equation}
      P(x)^{(i)} = \frac{\text{exp}(q_{\theta}(x)^{(i)} / \tau)}{\sum_{k = 1}^{K} \text{exp}(q_{\theta}(x)^{(k)} / \tau)}
      \label{eq:softmax_temp}
\end{equation}
where $P$ is the probability distribution with a given input $x$, $\tau$ is a temperature parameter which can control the sharpness of $P$. A larger $\tau$ smoothes $P$, while a smaller $\tau$ encourages the sharper output distribution. For example, if we set $\tau = 0$, then Eq.\eqref{eq:softmax_temp} is equivalent to the One-hot encoding. Here, we adopt different temperature parameters in the teacher and student networks, denoted as $\tau_{t}$ and $\tau_{s}$, respectively. In addition, we require that $\tau_{t}$ must be lower than $\tau_{s}$, and this design is called "sharpening." Sharpening is to avoid the mode collapse in the training process, which will be discussed in detail later. Moreover, as shown in Fig.\ref{fig:DINO}, compared with the student network, another design, called "centering", has been added to the teacher network. The specific approach is to add a bias item $c \in \mathbb{R}^{K}$ to the output of the teacher network and update it among different batches based on the Exponential Moving Average (EMA) approach in the training process, which can be described as Eq.\eqref{eq:centering}
\begin{equation}
      \begin{aligned}
            q_{\theta_{t}} (x) & \leftarrow q_{\theta_{t}} (x) + c                                                  \\
            c                  & \leftarrow m_{c}c + (1 - m_{c}) \frac{1}{B} \sum_{i = 1}^{B} q_{\theta_{t}} (x[i]) \\
      \end{aligned}
      \label{eq:centering}
\end{equation}
where $m_{c} \in [0, 1]$ is the momentum parameter in the centering, $B$ is batch size. In the same way as sharpening, centering can also avoid the mode collapse, whose impact will be presented later. Based on the above statement, the goal of DINO is to minimize the loss function shown in Eq.\eqref{eq:DINO_loss}
\begin{equation}
      \mathcal{L}_{\theta_{s}, \theta_{t}} = \sum_{x_{t} \in X^{g}} \sum_{\substack{x_{s} \in X' \\ x_{s} \neq x_{t}}} \mathcal{L}_{ce} (P_{s} (x_{s}), P_{t} (x_{t}))
      \label{eq:DINO_loss}
\end{equation}
where $\mathcal{L}_{ce}$ is the cross-entropy between the $P_{s}$ and $P_{t}$, and $\mathcal{L}_{ce} (a, b) = -a \text{ln} b$. We call the $\mathcal{L}_{\theta_{s}, \theta_{t}}$ as "target entropy".

Finally, we explain the parameter updating process of the teacher and student networks. As shown in Fig.\ref{fig:DINO}, $\theta_{s}$ is iteratively updated by the backpropagation. Furthermore, when constructing the computation graph, we apply the stop-gradient in the teacher network, that is, $\theta_{t}$ is not updated based on the loss function's gradient. In each training epoch, the iteration form of $\theta_{t}$ is similar to Eq.\eqref{eq:centering}, EMA approach can be described as
\begin{equation}
      \theta_{t} \leftarrow m \theta_{t} + (1 - m) \theta_{s}
      \label{eq:teacher_network}
\end{equation}
where $m \in [0, 1]$ is the momentum parameter. The process of DINO is shown in Algorithm.\ref{ag:DINO}. As the training process goes on, $\theta_{t}$ and $\theta_{s}$ keep iterating and updating. When the training is over, all the encoder parameters in the teacher network and student networks are frozen, and the feature representation $q_{\theta} (x)$ of the given TFM can be obtained.

\begin{algorithm}[h]
      \caption{The process of DINO algorithm}
      \label{ag:DINO}
      \hspace*{0.02in}{\bf Input:}
      training TMF dataset (without labels), Dataloader with batch size $B$ \\
      \hspace*{0.02in}{\bf Initialization:}
      teacher network $q_{\theta_{t}}$, student network $q_{\theta_{s}}$, bias term in centering $c$ \\
      \hspace*{0.02in}{\bf Define:}
      data augmentation strategy for global views $Aug_{g}(\cdot)$, data augmentation strategy for local views $Aug_{l}(\cdot)$

      \begin{algorithmic}[1]
            \FOR{$epoch$ in range(max\_epoch)}
            \FOR{$x$ in Dataloader}
            \STATE{Obtain the global crops: $X^{g} = Aug_{g}(x)$}
            \STATE{Obtain the local crops: $X^{l} = Aug_{l}(x)$, and let $X' = X^{g} \cup X^{l}$}
            \STATE{Centering operation: $q_{\theta_{t}} (x) \leftarrow q_{\theta_{t}} (x) + c$}
            \STATE{Calculate the target entropy: $\mathcal{L}_{\theta_{s}, \theta_{t}} = \sum_{x_{t} \in X^{g}} \sum_{\substack{x_{s} \in X' \\ x_{s} \neq x_{t}}} \mathcal{L}_{ce} (P_{s} (x_{s}), P_{t} (x_{t}))$}
            \STATE{Apply the stop-gradient in the teacher network: $\text{stop-gradient} (q_{\theta_{t}})$}
            \STATE{Backpropagation for the student network: $\theta_{s} \leftarrow \text{optimizer}(\theta_{s}, \nabla \mathcal{L}_{\theta_{s}, \theta_{t}})$}
            \STATE{Update the teacher network by EMA: $\theta_{t} \leftarrow m \theta_{t} + (1 - m) \theta_{s}$}
            \STATE{Update the bias term by EMA: $c \leftarrow m_{c}c + (1 - m_{c}) \frac{1}{B} \sum_{i = 1}^{B} q_{\theta_{t}} (x[i])$}
            \ENDFOR
            \ENDFOR
      \end{algorithmic}
\end{algorithm}

\subsection{Overview}

\begin{figure}[htbp]
      \centering
      \includegraphics[width = 0.99\textwidth]{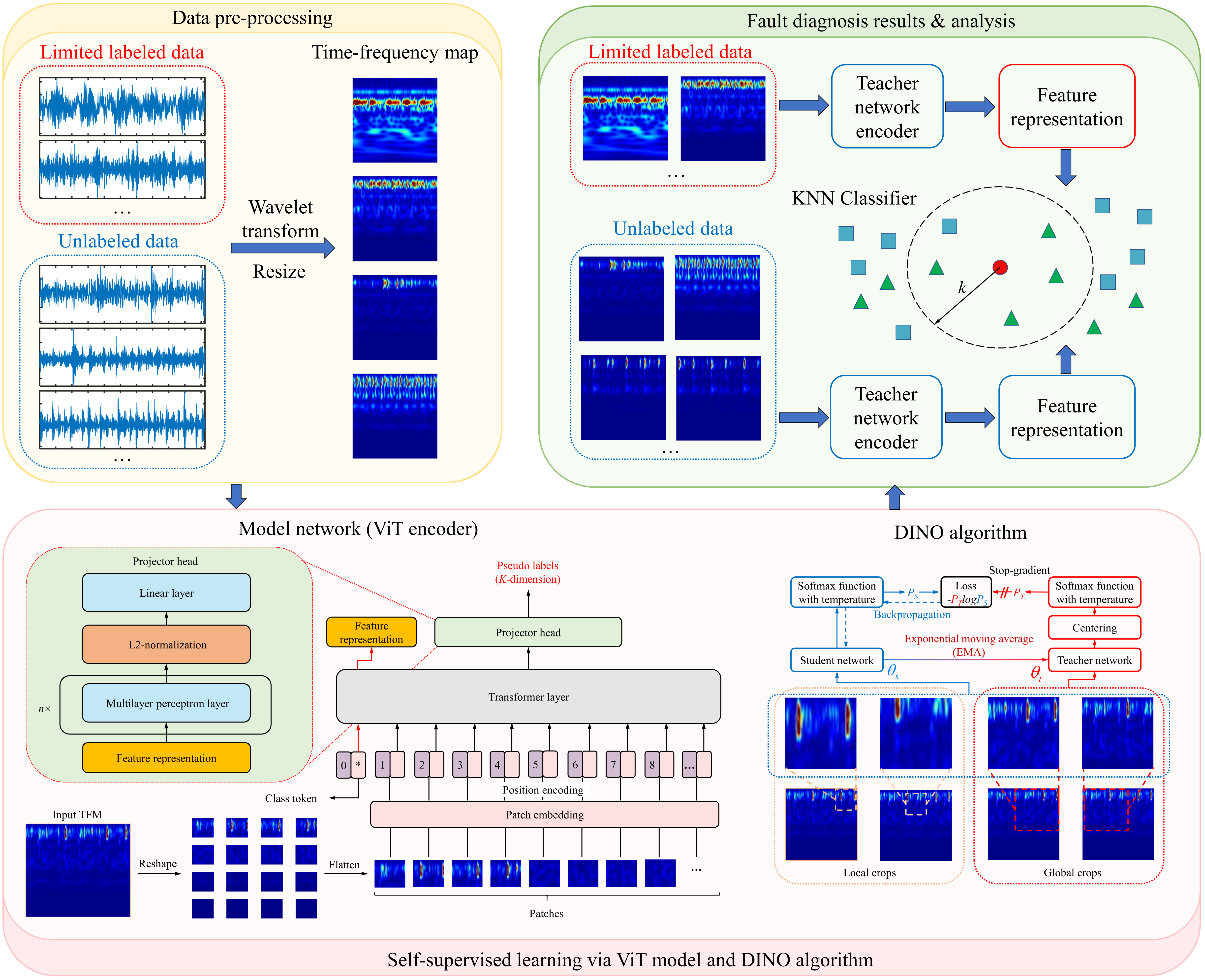}
      \caption{The overall framework of the proposed method.}
      \label{fig:method_framework}
\end{figure}

We propose a new intelligent fault diagnosis method for the rolling bearings based on the DINO algorithm and ViT model to solve the fault diagnosis problem under the limited labeled data condition. The process framework is illustrated in Fig.\ref{fig:method_framework}, and its specific stages can be described as follows

1) Converting the original vibration signals into TFMs through the WT and resize operation.

2) Constructing the model network based on the ViT.

3) Combining the limited labeled and unlabeled data and conducting the self-supervised learning via the ViT model and DINO algorithm.

4) Freezing and saving the model networks' parameters. Adopting the encoder part of the teacher network to extract the feature representation from the input TFMs.

5) Outputting the fault diagnosis results based on the limited labeled data and K-nearest neighbor (KNN) classifier.

\section{Experimental setup}

\subsection{Datasets description}

In this paper, we adopt the CWRU dataset \cite{CWRUDataset}, which is collected by the Bearing Data Center of Case Western Reserve University (CWRU), and the XJTU dataset \cite{8576668}, which is provided by the Xi'an Jiaotong University (XJTU), to train the model network and evaluate the effectiveness of the proposed method. Both are publicly available rolling bearing fault datasets widely used by many researchers. The following is a brief Introduction to the datasets.

\begin{figure}[h]
      \centering
      \includegraphics[width = 0.69\textwidth]{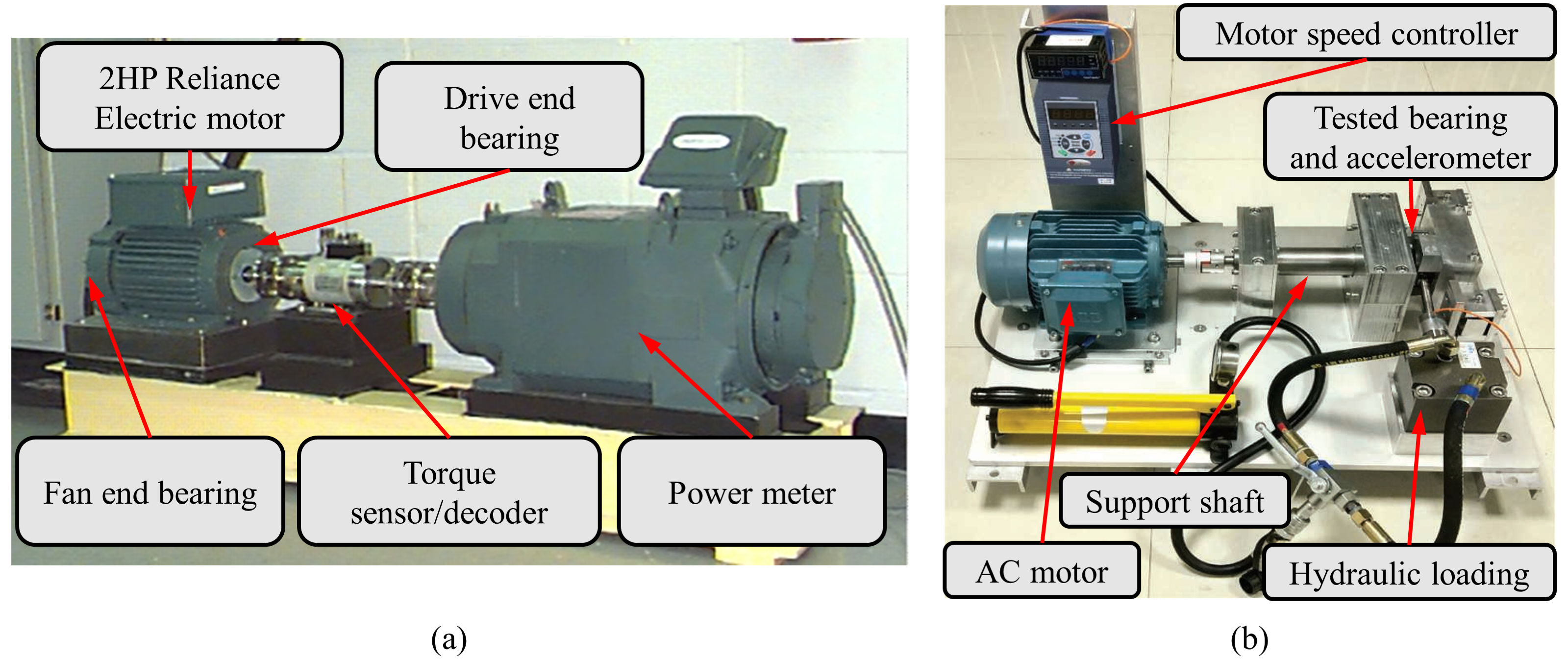}
      \caption{Test rig of the datasets. (a) CWRU dataset. (b) XJTU dataset.}
      \label{fig:exp_device}
\end{figure}

1) The CWRU dataset is a rolling bearing prefabricated fault dataset. The dataset is composed of multivariate vibration signals generated by a bearing test-rig, as presented in Fig.\ref{fig:exp_device}(a). The vibration signals are measured by the acceleration sensor with 12kHz sampling rates. We adopt the bearing data from the drive end of the motor (bearing type 6205-2RS JEM SKF) in this study, and three fault types are processed with the Electro-Discharge Machining (EDM), including inner race fault (IR), outer race fault (OR), and rolling ball fault (RB). Besides, various fault diameters ranging from 7 inches to 21 mils are introduced in each fault type, separately. In summary, the CWRU dataset contains ten failure modes (including the normal condition, denoted as NC). Then, based on the resampling method, the specific composition of the CWRU dataset in this paper is shown in Tab.\ref{tb:datasets}, where only 1\% limited labeled data is involved.

2) The XJTU dataset comprises complete run-to-failure data of 15 rolling element bearings (same type, LDK UER204) acquired by conducting many accelerated degradation experiments, whose experimental device is shown in Fig.\ref{fig:exp_device}(b). Vibration signals of the tested bearings are obtained by two acceleration sensors in horizontal and vertical directions with 25.6kHz sampling rates. In this paper, the recorded data from Bearing 3\_1, Bearing 3\_2, and Bearing 2\_3 are chosen as the analysis data, which contains four fault modes: Outer race (OR), Inner race, ball, cage and outer race (IBCO), Inner race (IR), and Cage. Then, similar to the CWRU dataset, the XJTU dataset after resampling is shown in Tab.\ref{tb:datasets}, which also contains only 1\% labeled data. 

\begin{table}[h]
      \caption{Composition of the datasets}\label{tb:datasets}
      \begin{tabular*}{1.0\textwidth}{@{}LLLLL@{}}
            \toprule
            \multirow{2}*{Dataset} & \multirow{2}*{Fault modes} & \multirow{2}*{Sample size} & \multirow{2}*{\makecell[l]{The number of \\ labeled data}} & \multirow{2}*{\makecell[l]{The number of \\ unlabeled data}} \\
            ~ & ~ & ~ & ~ & ~ \\
            \midrule
            \multirow{2}*{CWRU} & \multirow{2}*{\makecell[l]{NC, IR007 to IR021 \\ OR007 to OR021, RB007 to RB021}} & \multirow{2}*{1000 $\times$ 10} & \multirow{2}*{10 $\times$ 10} & \multirow{2}*{990 $\times$ 10} \\
            ~ & ~ & ~ & ~ & ~ \\
            XJTU & OR, IBCO, IR, Cage & 2000 $\times$ 4 & 20 $\times$ 4 & 1980 $\times$ 4 \\
            \bottomrule
      \end{tabular*}
\end{table}

\subsection{Parameters setup and other details}

During the training process, the structural parameters of the model network adopted in this study are shown in Tab.\ref{tb:model_network_para}. The backbone of the encoder follows a lightweight ViT architecture, where the patch size of the input TFM $P$ is 16, the embedding dimension $d$ is 192, the number of $head$ in Eq.\eqref{eq:multi_head_attn} is 3, the dimension of the keys and values is 64, the embedding dimension of the nonlinear projection layer in the MLP block $d_{MLP}$ satisfies $d_{MLP} = 4 d$, and the number of stacked Transformer basic blocks $depth$ is 12. Besides, for the projector head, the dimension of the pseudo labels $K$ is 1024, and there are 3-layer MLP, where the dimension of the hidden and bottleneck layer $d_{MLP}^{head}$ is (2048, 2048, 256), respectively. Then, the parameters of the DINO algorithm are shown in Tab.\ref{tb:DINO_para}. The momentum parameter of the teacher network $m$ in Eq.\eqref{eq:teacher_network} follows the cosine schedule from 0.996 to 1 during the training process. The momentum parameter in the centering is 0.9. The temperature of the teacher and student networks ($\tau_{t}$ and $\tau_{s}$) are 0.04 and 0.1, separately. As described previously, $\tau_{t}$ should be lower than $\tau_{s}$ for sharpening. Moreover, the scale parameter $s$ is 0.4, which means that the scale range of the local views is $[0.05, s]$, and the scale range of the global views is $[s, 1]$. The number of local crops $N$ is 8. The above parameters are adjustable hyperparameters, and we will discuss their impact on fault diagnosis performance in detail later. 

\begin{table}[h]
      \caption{Structural parameters of the model network.}\label{tb:model_network_para}
      \begin{tabular*}{1.0\textwidth}{@{}LLLLLLLLLLL@{}}
            \toprule
            \multirow{2}*{Name} & \multirow{2}*{Input size} & \multicolumn{6}{c}{Encoder} & \multicolumn{3}{c}{Projector head} \\
            ~ & ~ & $P$ & $d$ & $h$ & $d_{K} = d_{V} $ & $d_{MLP}$ & $depth$ & $K$ & $n$ & $d_{MLP}^{head}$ \\
            \midrule
            Value & 224 $\times$ 224 $\times$ 3 & 16 & 192 & 3 & 64 & 4 $\times$ 192 & 12 & 1024 & 3 & (2048, 2048, 256) \\
            \bottomrule
      \end{tabular*}
\end{table}

\begin{table}[h]
      \caption{Parameters of the DINO algorithm.}\label{tb:DINO_para}
      \begin{tabular*}{1.0\textwidth}{@{}LLLLLLL@{}}
            \toprule
            Name & $m$ & $m_{c}$ & $\tau_{t}$ & $\tau_{s}$ & $[0.05, s], [s, 1], s$ & $N$ \\
            \midrule
            Value & 0.996 $\rightarrow$ 1 & 0.9 & 0.04 & 0.1 & 0.4 & 8 \\
            \bottomrule
      \end{tabular*}
\end{table}

Finally, we adopt the Adam optimizer to minimize the target entropy in Eq.\eqref{eq:DINO_loss}, and set the weight decay equal to 0.04 to introduce the regularization. The batch size $B$ is 64, and the max training epoch is set to 100. Then, we use the warm-up strategy combined with cosine schedule to adjust the learning rate. At the first ten training epochs, the learning rate increases linearly from $1 \times 10^{-6}$ to $1.25 \times 10^{-4}$, and the subsequent learning rate follows the cosine schedule. The hardware environment is Ryzen 3995WX, NVIDIA RTX 3090, and we adopt Python 3.8, Pytorch 1.8.1, and CUDA 10.2 for the deep learning framework. 

\section{Experimental results and discussions}

\subsection{Mode collapse problem}

In self-supervised learning, mode collapse is a crucial problem. Mode collapse refers to the situation in which the network gradually converges to trivial solutions due to the abnormal training process. For example, the trained adversarial generation network's generator can only generate one kind of image. As shown in Fig.\ref{fig:model_network} and Fig.\ref{fig:DINO}, DINO's goal is to make the $K$-dimensional pseudo labels of the student network match the teacher network. Based on this, there are two forms of the collapse in the proposed method: 1) the outputs of the teacher and student networks are evenly distributed in each dimension (namely over-alignment), that is, the probability value of each pseudo label is $\frac{1}{K}$. 2) the output of the teacher and student networks is 1 in one dimension and 0 in all others (namely over-uniformity), such as $[0, 1, 0, 0, ..., 0]$. Then, as described previously, we introduce centering and sharpening to avoid collapse, and we will study their respective roles in this section. Firstly, the cross entropy in Eq.\eqref{eq:DINO_loss} can be decomposed into Kullback-Leibler (KL) divergence and entropy
\begin{equation}
      \sum_{x_{t} \in X^{g}} \sum_{\substack{x_{s} \in X' \\ x_{s} \neq x_{t}}} \mathcal{L}_{ce} (P_{s} (x_{s}), P_{t} (x_{t})) = \sum_{x_{t} \in X^{g}} \sum_{\substack{x_{s} \in X' \\ x_{s} \neq x_{t}}} D_{KL} (P_{s} (x_{s}) | P_{t} (x_{t})) + \sum_{x_{t} \in X^{g}} h(P_{t} (x_{t}))
      \label{eq:KL_div}
\end{equation}
where $D_{KL}$ is the KL divergence between $P_{t}$ and $P_{s}$, and $h$ is the entropy of the teacher network's output $P_{t}$. $D_{KL}$ can calculate the match between $P_{t}$ and $P_{s}$. If $P_{t}$ and $P_{s}$ are identical, then $D_{KL}$ equals zero, which means the mode collapse. Besides, $h$ can measure the uncertainty of $P_{t}$, which can be given as
\begin{equation}
      h(P_{t}) = \sum_{i = 1}^{K} -P_{t}^{(i)} ln P_{t}^{(i)}
      \label{eq:entropy}
\end{equation}
where $P_{t}^{(i)}$ is $i$th element in $P_{t}$. As mentioned above, the value of $D_{KL}$ can be used to judge whether the mode collapse occurs. Furthermore, we can determine the form of the mode collapse based on the value of $h$. It can be seen from Eq.\eqref{eq:entropy} that $h$ will converge to $\text{ln} K$ for over-uniformity case, and will converge to zero for over-alignment case. 

\begin{figure}[h]
      \centering
      \includegraphics[width = 0.95\textwidth]{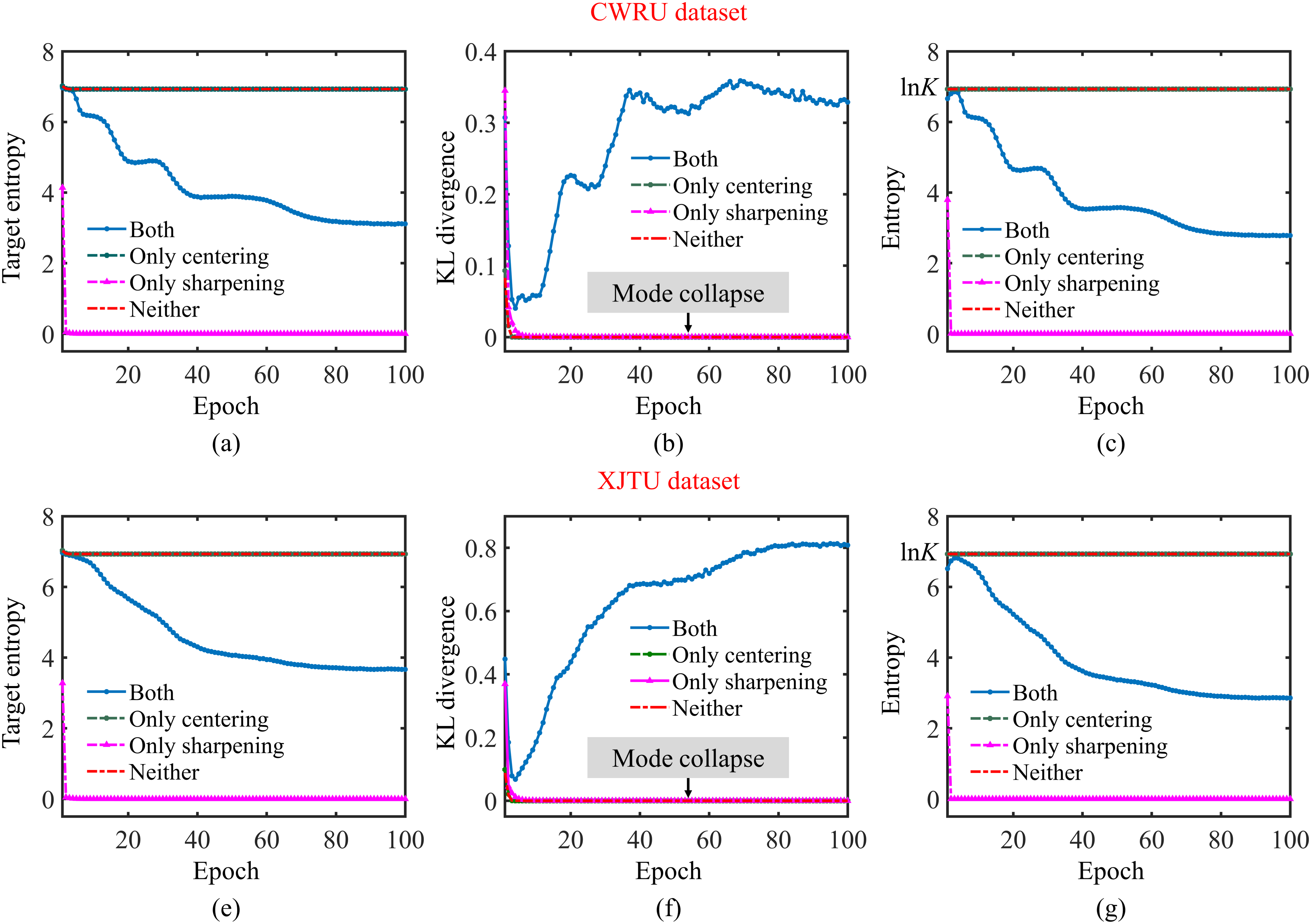}
      \caption{Mode collapse study in four designs under the different datasets. (a)-(c) CWRU dataset. (e)-(g) XJTU dataset. (a) and (e) target entropy. (b) and (f) KL divergence. (c) and (g) entropy.}
      \label{fig:mode_collapse}
\end{figure}

Subsequently, we present the evolution of $\mathcal{L}_{\theta_{s}, \theta_{t}}$, $D_{KL}$ and $h$ under different datasets and various designs in Fig.\ref{fig:mode_collapse}. To analyze the respective roles of the centering and sharpening, we set up four different designs: 1) The output of the teacher network goes through both centering and sharpening operations (namely Both), as shown in Fig.\ref{fig:DINO}. 2) There is only centering in the DINO framework. 3) There is only sharpening in the DINO framework. 4) The output of the teacher network is identical to the student network without any additional modifications (namely Neither). As shown in Fig.\ref{fig:mode_collapse}(a) and (e), in the case of only centering, the target entropy will stay at an enormous value during the entire training process. Conversely, in the case of only sharpening, the target entropy converges rapidly to zero. Both cases indicate that the model networks have not been trained normally. Further, it can be seen from Fig.\ref{fig:mode_collapse}(b) and (f) that the KL divergence between the teacher and student outputs degrades to zero promptly in all cases except Both, which means $P_{t} = P_{s}$, indicating the mode collapse. However, the entropy converges to different values under the diverse designs. It can be seen from the data in Fig.\ref{fig:mode_collapse}(c) and (g) that if there is only centering operation, the entropy of the teacher network output $h$ will converge to $\text{ln} K$, indicating the over-uniformity. And $h$ will converge to 0 if there is only sharpening in the DINO framework, which means the over-alignment case. In the absence of centering and sharpening, the result is the same as that of only centering, and over-uniformity occurs. Finally, from the blue lines in Fig.\ref{fig:mode_collapse} we can see that $\mathcal{L}_{\theta_{s}, \theta_{t}}$ and $h$ show a tendency to decline gradually during training, and $D_{KL}$ do not converge to 0, indicating that there is no mode collapse. These results suggest that the teacher network and the student network can carry out normal training only in the case of Both. 

In summary, for the informants in this section, it can be observed that the centering and sharpening operations are complementary in avoiding the mode collapse. Missing either operation will cause the mode collapse in the proposed method. Centering can inhibit over-alignment, but encourage over-uniformity. Sharpening has the opposite effect, which can avoid over-uniformity but promote over-alignment. In addition, mode collapse will also occur if there is no centering or sharpening. Only by applying both operations simultaneously can the model network be appropriately trained. 

\subsection{Fault diagnosis results based on the proposed method}

Based on the DINO algorithm, the model networks are trained on the CWRU and XJTU datasets without any labels. The parameters in the encoder part of the teacher network are frozen, and we can use it to extract the feature representation of the input TFM. The extracted feature representation is called feature vectors. Then, to diagnosis the input TFM, we calculate its feature vectors and compare it against the labeled data. Adopting the above process, we can make fault diagnose for many unlabeled data with the KNN classifier, even if the labeled data is minimal. Tab.\ref{tb:results_acc} provides the fault diagnosis results obtained by our proposed method. We select multiple the number of nearest neighbors (denoted as $N_{k}$), ranging from 10 to 100, to optimize the classifier parameters. In addition, to enhance the robustness of the classifier to $N_{k}$, we introduce a temperature parameter $\tau_{k}$ as a divisor when calculating the similarity, similar to Eq.\eqref{eq:softmax_temp}. The effect of $\tau_{k}$ is also set out in Tab.\ref{tb:results_acc}. We find that $N_{k}$ has a significant impact on the fault diagnosis accuracy of the KNN classifier if without $\tau_{k}$. On the CWRU dataset, the accuracy of the classifier is 92.65\% with $N_{k} = 10$, but when $N_{k}$ equals 40, the accuracy drops to 79.19\%. Analogously, the fault diagnosis accuracy of KNN is 83.08\% with $N_{k} = 10$ and 74.07\% with $N_{k} = 40$ on the XJTU dataset. These results show that the fault diagnosis performance of the classifier will be sensitive to $N_{k}$ if there is no temperature parameter. Furthermore, the first and third rows in Tab.\ref{tb:results_acc} illustrate that $\tau_{k}$ can obviously improve the robustness of the classifier to $N_{k}$, and increase the fault diagnosis accuracy. Taking the CWRU dataset as an example, for the different values of $N_{k}$, the top and minimum accuracy of the classifier are 95.27\% and 93.44\%, and their fluctuation range is significantly reduced compared with the data without $\tau_{s}$. Then, by applying $\tau_{s}$, the accuracy of KNN on the CWRU and XJTU datasets achieve 95.27\% and 89.33\% with $N_{k} = 10$, respectively. This is a brilliant fault diagnosis result under the limited labeled data condition. Moreover, it should be pointed out that the feature vectors extracted by the trained encoder do not undergo any additional fine-tuning, and great fault diagnosis accuracy shows that feature vectors possess good generalization ability. 

\begin{table}[h]
      \caption{Fault diagnosis results with only 1\% labeled data (unit: \%).}\label{tb:results_acc}
      \begin{tabular*}{1.0\textwidth}{@{}LLLLLLLLLLL@{}}
            \toprule
            Dataset $\backslash$ $N_{k}$ & 10 & 20 & 30 & 40 & 50 & 60 & 70 & 80 & 90 & 100 \\
            \midrule
            CWRU with $\tau_{k}$ & \textbf{95.27} & 93.69 & 93.44 & 93.46 & 93.47 & 93.47 & 93.74 & 93.47 & 93.47 & 93.47 \\
            CWRU without $\tau_{k}$ & 92.65 & 81.42 & 79.22 & 79.19 & 79.49 & 80.05 & 80.58 & 80.73 & 81.22 & 81.22 \\
            XJTU with $\tau_{k}$ & \textbf{89.33} & 88.32 & 88.40 & 88.37 & 88.37 & 88.37 & 88.37 & 88.37 & None & None \\
            XJTU without $\tau_{k}$ & 83.08 & 75.74 & 74.44 & 74.07 & 75.25 & 75.34 & 75.52 & 75.77 & None & None \\
            \bottomrule
      \end{tabular*}
\end{table}

\begin{figure}[h]
      \centering
      \includegraphics[width = 0.62\textwidth]{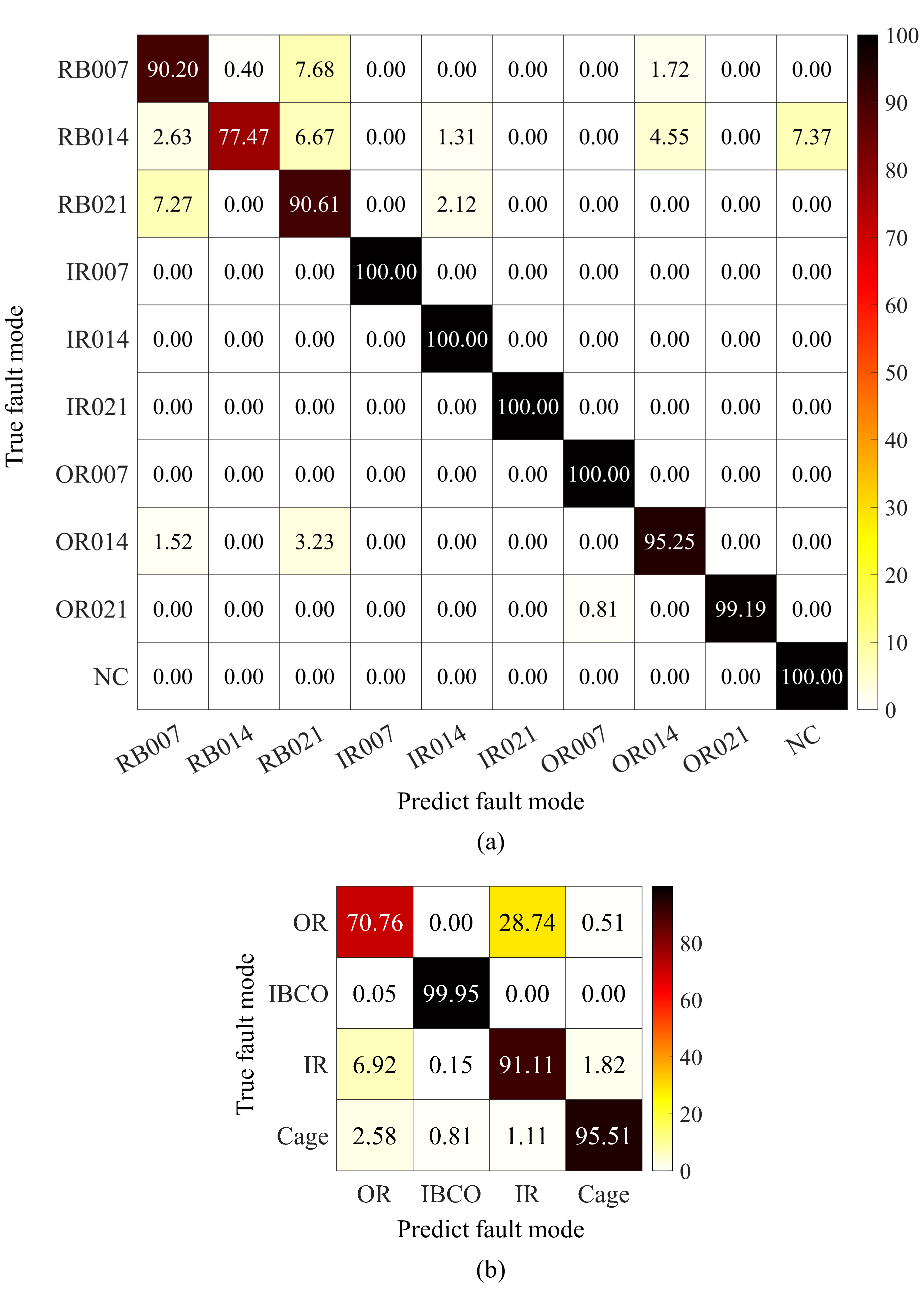}
      \caption{Confusion matrix results. (a) CWRU dataset. (b) XJTU dataset.}
      \label{fig:confusion}
\end{figure}

To analyze the detailed fault diagnosis results, Fig.\ref{fig:confusion} provides the confusion matrices of the proposed method on the CWRU and XJTU datasets, where rows represent the actual fault mode and columns is the predicted fault mode. Specifically, 
on the CWRU dataset, promising fault diagnosis accuracy can be obtained in most fault modes, and the misidentification results mainly focus on the class RB014. This result may be explained by the fact that the features of rolling ball fault condition is general weak. However, we can still obtain higher than 77\% identification accuracy with only 1\% labeled data. In contrast, due to the earlier fault stage of the rolling bearing during the run-to-failure process, the proposed method achieves relatively lower diagnosis accuracy on the XJTU dataset. About 89.33\% testing accuracy with the unlabeled data can be acquired, and 28.74\% OR samples are misclassified as IR. This result indicates that the outer race fault in early stage may be a challenging fault mode. But, over 70\% accuracy for OR can be still achieved. That shows that the proposed method has satisfactory fault diagnosis performance in the case of limited labeled data. 

Overall, through the results of fault diagnosis accuracy and confusion matrices under the different datasets, the effectiveness of the proposed method is validated in this section. Feature vectors with strong generalization ability can be acquired utilizing the model networks and DINO algorithm, and high diagnosis accuracy is obtained with only 1\% labeled data. 

\subsection{Influence of the hyperparameters}

In this section, we will investigate the influence of the hyperparameters in the proposed method, including the encoder form, pseudo labels dimension $K$, temperature parameters in the teacher and student networks ($\tau_{t}$ and $\tau_{s}$), scale parameter $s$, momentum parameter in centering $m_{c}$, and the number of local crops $N$. The impact of the hyperparameters is evaluated from two aspects: accuracy and peak memory when running. The accuracy directly reflects the fault diagnosis performance, while the peak memory can represent the model scale and computational complexity. These results are summarized in Tab.\ref{tb:Hyperparameters}. Firstly, the result in group A shows that we can obtain a higher accuracy based on the feature vectors extracted by the teacher network. A possible explanation for this might be that the outputs of the teacher network are sharper ($\tau_{t} < \tau_{s}$), which helps the model learn features with more generalization ability. Next, as one of the critical hyperparameters in the proposed method, the effects of the pseudo labels dimension $K$ are discussed in group B in Tab.\ref{tb:Hyperparameters}. One unanticipated finding is that a larger $K$ may not imply better performance. For example, the fault diagnosis accuracy of $K = 64$ is higher than that of $K = 512$. This result suggests that the performance of the proposed method may be further improved by fine-tuning $K$. Within the range that we set in this study, the accuracy is highest when $K$ equals 1024. Meanwhile, the data in group B also show that the size of $K$ has no significant impact on the peak memory. This is because $K$ is associated only with the last layer of the projector head, which is a negligible value compared to the entire model network. 

Besides, the relationship between the temperature parameter of teacher network and the testing accuracy is presented in group C in Tab.\ref{tb:Hyperparameters}. We observe that the model is susceptible to $\tau_{t}$, and inappropriate $\tau_{t}$ will significantly reduce the model performance, even leading to a failed training process. Too small $\tau_{t}$ will make the teacher output too sharp, which causes the network easily fall into the local minimum, resulting in the decline of diagnosis accuracy. Nevertheless, as mentioned in Section 4.1, we need sharpening to avoid the mode collapse, so a too large $\tau_{t}$ is also unbefitting. Actually, through the experiment, we find that when the teacher temperature is higher than 0.06 (such as 0.8), the entropy of teacher output consistently converges to $\text{ln} K$, indicating over-uniformity. Therefore, the fault diagnosis performance is inferior with an overlarge $\tau_{t}$. 

Next, group D in Tab.\ref{tb:Hyperparameters} compares the testing accuracy and peak memory of the proposed method under diverse student temperature $\tau_{s}$. As described previously, the design of sharpening demands that $\tau_{t}$ must be lower than $\tau_{s}$ to avoid the mode collapse, further confirmed by the result in group D. If $\tau_{t}$ is higher than or close to $\tau_{s}$, such as 0.03 or 0.05, sharpening will fail, and the poor diagnosis performance will be obtained. In our experiment, the student temperature below 0.05 is not recommended. In addition, similar to $\tau_{t}$, oversize $\tau_{s}$ is able to reduce the fault diagnosis accuracy. Hence, a moderate $\tau_{s}$ is suggested in practical implementations. 

Scale range $s$ is also an essential parameter that affects the fault diagnosis performance of the proposed method, and the testing results with different $s$ are presented in group E. A general pattern summarized from group E is that higher diagnosis accuracy can be acquired with a medium size of $s$. A lower $s$ means smaller areas of the original TFM adopted to generate the global and local crops, which encourages the model networks to concentrate on the local features of the input. On the contrary, a larger $s$ helps our proposed method to learn the global features of the TFM. Both local and global features are necessary for the fault diagnosis problem, so a moderate $s$ should be selected to balance their effects. In our study, the highest accuracy can be achieved with $s = 0.4$. Moreover, since the local and global crops are resized to a fixed size, $s$ does not influence the computational complexity. 

Per our earlier discussion, there are two additional designs in the teacher output to prevent the mode collapse: sharpening and centering. $\tau_{t}$ and $\tau_{s}$ are related to sharpening, and their effects have been investigated in group C and D. $m_{c}$ is the momentum parameter in centering, and its impact is provided in group F. It is somewhat surprising that, compared with the temperature parameters, the convergence of the proposed method is robust to a wide range of $m_{c}$. We do not notice the over-uniformity or over-alignment with $m_{c} \in [0, 0.99]$ in our trials. The model network only occurs the model collapse when the update of bias term in centering is too slow, such as $m_{c} = 0.999$. Simultaneously, the maximum testing accuracy can be procured with $m_{c}$ equals 0.9 in our datasets. 

Finally, the influence of the number of local crops $N$ is shown in group G in Tab.\ref{tb:Hyperparameters}. A universal display pattern is observed that a larger $N$ can improve the diagnosis performance. Nonetheless, the computational complexity of the proposed method also increases with the expansion of $N$, so we should weigh it according to the specific hardware environment in the actual application. 

\begin{table}[h]
      \caption{Influence of the hyperparameters}\label{tb:Hyperparameters}
      \begin{tabular*}{1.0\textwidth}{@{}LLLLLLLLLL@{}}
            \toprule
            \multirow{2}*{Group} & \multicolumn{7}{c}{Hyperparameters} & \multirow{2}*{Accuracy} & \multirow{2}*{Peak memory} \\
            ~ & Encoder form & $K$ & $\tau_{t}$ & $\tau_{s}$ & $s$ & $m_{c}$ & $N$ & ~ & ~ \\
            \midrule
            Baseline & Teacher & 1024 & 0.04 & 0.1 & 0.4 & 0.9 & 8 & \textbf{95.27\%} & 6.34G \\
            \midrule
            A & Student & ~ & ~ & ~ & ~ & ~ & ~ & 94.34\% & 6.34G \\
            \midrule
            \multirow{5}*{B} & ~ & 32 & ~ & ~ & ~ & ~ & ~ & 92.65\% & 6.33G \\
            ~ & ~ & 64 & ~ & ~ & ~ & ~ & ~ & 94.24\% & 6.33G \\
            ~ & ~ & 128 & ~ & ~ & ~ & ~ & ~ & 93.72\% & 6.33G \\
            ~ & ~ & 256 & ~ & ~ & ~ & ~ & ~ & 92.10\% & 6.33G \\
            ~ & ~ & 512 & ~ & ~ & ~ & ~ & ~ & 90.86\% & 6.33G \\
            \midrule
            \multirow{3}*{C} & ~ & ~ & 0.02 & ~ & ~ & ~ & ~ & 94.85\% & 6.34G \\
            ~ & ~ & ~ & 0.06 & ~ & ~ & ~ & ~ & 88.79\% & 6.34G \\
            ~ & ~ & ~ & 0.08 & ~ & ~ & ~ & ~ & 37.82\% & 6.34G \\
            \midrule
            \multirow{4}*{D} & ~ & ~ & ~ & 0.03 & ~ & ~ & ~ & 51.06\% & 6.34G \\
            ~ & ~ & ~ & ~ & 0.05 & ~ & ~ & ~ & 52.00\% & 6.34G \\
            ~ & ~ & ~ & ~ & 0.2 & ~ & ~ & ~ & 94.24\% & 6.34G \\
            ~ & ~ & ~ & ~ & 0.4 & ~ & ~ & ~ & 86.72\% & 6.34G \\
            \midrule
            \multirow{5}*{E} & ~ & ~ & ~ & ~ & 0.08 & ~ & ~ & 87.91\% & 6.34G \\
            ~ & ~ & ~ & ~ & ~ & 0.16 & ~ & ~ & 92.97\% & 6.34G \\
            ~ & ~ & ~ & ~ & ~ & 0.24 & ~ & ~ & 93.82\% & 6.34G \\
            ~ & ~ & ~ & ~ & ~ & 0.32 & ~ & ~ & 94.08\% & 6.34G \\
            ~ & ~ & ~ & ~ & ~ & 0.48 & ~ & ~ & 92.09\% & 6.34G \\
            \midrule
            \multirow{3}*{F} & ~ & ~ & ~ & ~ & ~ & 0 & ~ & 91.32\% & 6.34G \\
            ~ & ~ & ~ & ~ & ~ & ~ & 0.99 & ~ & 92.27\% & 6.34G \\
            ~ & ~ & ~ & ~ & ~ & ~ & 0.999 & ~ & 63.60\% & 6.34G \\
            \midrule
            \multirow{4}*{G} & ~ & ~ & ~ & ~ & ~ & ~ & 0 & 86.67\% & 4.27G \\
            ~ & ~ & ~ & ~ & ~ & ~ & ~ & 2 & 90.70\% & 4.69G \\
            ~ & ~ & ~ & ~ & ~ & ~ & ~ & 4 & 91.58\% & 5.14G \\
            ~ & ~ & ~ & ~ & ~ & ~ & ~ & 6 & 93.71\% & 5.78G \\
            \bottomrule
      \end{tabular*}
\end{table}

\subsection{Attention maps visualization}

Great interpretability is a significant advantage of the attention mechanism. This section employs attention maps (denoted as AM) visualization to explore the feature representation process qualitatively. Here, three dissimilar forms of AM are introduced: 1) Class token AM (CAM), which is calculated by the class token and embedding sequence on the heads of the last Transformer basic block in the model network. CAM can represent the importance of each patch in the input TFM to the feature representation. 2) Threshold class token AM (TAM). We visualize the marked patches obtained by thresholding the CAM to keep 90\% of the attention, illustrating the areas that primarily affect the feature vectors. 3) Embedding sequence AM (EAM), computing from the embedding sequence in the last Transformer basic block. Note that there are three $head$ in the MSA, so the EAM also can be described in three different subspaces, denoted as $\text{EAM}_1$, $\text{EAM}_2$, and $\text{EAM}_3$ respectively. 

\begin{figure}[h]
      \centering
      \includegraphics[width = 0.65\textwidth]{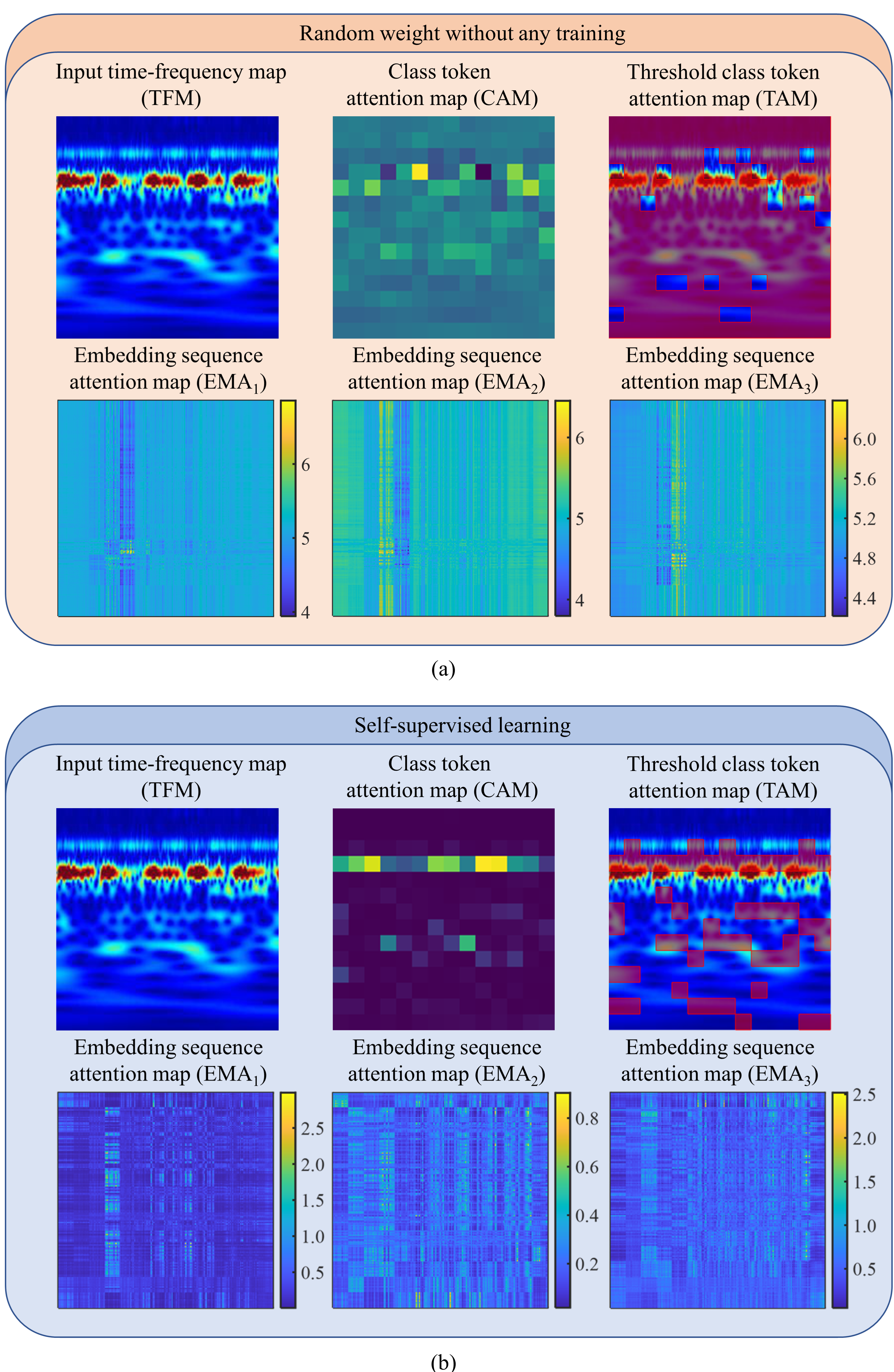}
      \caption{Attention maps visualization results. (a) Without training. (b) Self-supervised learning.}
      \label{fig:attn_map}
\end{figure}

The results of the attention maps visualization obtained from the untrained and trained teacher network are compared in Fig.\ref{fig:attn_map}. As shown in Fig.\ref{fig:attn_map}(a), for the network adopting the random weight without any training, no salient attention value is found in CAM, and marked patches in TAM almost cover the whole input region. Additionally, the attention values in EAM also present a consistent distribution form, indicating a poor feature extraction ability. In contrast, the teacher network trained by self-supervised learning demonstrates an entirely different pattern. As can be seen from Fig.\ref{fig:attn_map}(b), the prominent parts in TFM have more significant attention values in CAM, and the marked patches in TAM are also concentrated on the smaller areas. That is, the trained teacher network pays more attention to the patches where the amplitude in the TFM is more pronounced, which adheres to our intuition. Then, most of the attention values in EMA are centralized in a few patches, suggesting that the teacher network grasps the crucial information for fault diagnosis from the input TFM. 

Taken together, through the attention maps visualization, the feature representation process in our proposed method is analyzed. Based on self-supervised learning, the teacher network spontaneously learns the fault-specific features without any labels, and it pays more attention to the patches where the amplitude in the TFM is more obvious. 

\subsection{Feature vectors visualization}

In addition to the testing accuracy, the distribution form of feature vectors is also a meaningful indicator to evaluate the proposed fault diagnosis method. Therefore, we will focus on discussing the visualization results of feature vectors in this section. 

\begin{figure}[h]
      \centering
      \includegraphics[width = 0.62\textwidth]{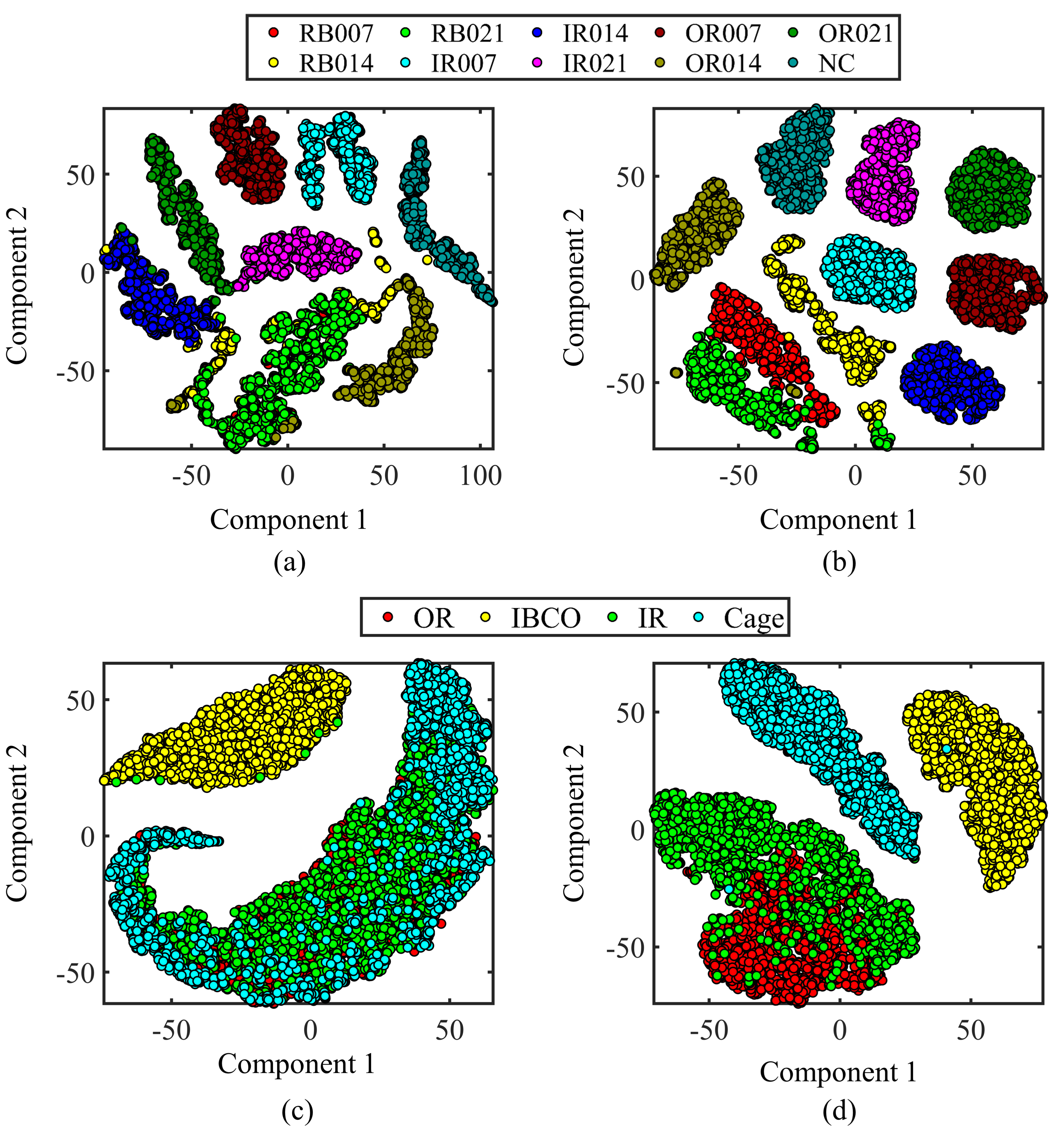}
      \caption{Feature vectors visualization results. (a) and (b) CWRU dataset. (c) and (d) XJTU dataset. (a) and (c) The input TFMs. (b) and (d) Feature vectors extracted by our proposed method}
      \label{fig:features_visualization}
\end{figure}

In this study, the dimension of the feature vectors extracted by the model network is 192, which is a higher-dimensional space that cannot be visualized directly. Here, we adopt the t-distributed Stochastic Neighbor Embedding (t-SNE) technique as a dimension reduction method. As a comparison, visualization results of the input TFMs are also necessary. However, it should be noted that the scale of the TFM is $224 \times 224 \times 3$, so processing it immediately with t-SNE will be very time-consuming and not effective. To address this issue, an approach combined the Principal Component Analysis (PCA) and t-SNE method is utilized. Firstly, PCA is used to decompose the first 192 principal components of the TFM and integrate them into a feature vector. Then, t-SNE is applied to reduce its dimension further and realize visualization. Based on the above analysis, Fig.\ref{fig:features_visualization} shows the feature vectors visualization results on the CWRU and XJTU datasets. A large amount of overlapping parts are observed for different fault modes in Fig.\ref{fig:features_visualization}(a) and (c), demonstrating that it is not feasible to use the input TFM for fault diagnosis directly. Then, turning now to the visualization results of the feature vectors obtained by our proposed method. As shown in Fig.\ref{fig:features_visualization}(b) and (d), most instances of the same fault mode are projected into the same region, and different fault modes are separated. That shows that the proposed method can extract the feature vectors with inter-class separability by exploring information in unlabeled data. However, we should point out that confusion still occurs between a few different samples due to the lack of supervised learning with labeled data, such as RB007 and RB021 in Fig.\ref{fig:features_visualization}(b), and OR and IR in Fig.\ref{fig:features_visualization}(d). These results are compatible with the above confusion matrix analysis results in Fig.\ref{fig:confusion}, revealing some more challenging fault diagnosis tasks. 

\subsection{Comparison with other methods}

To further verify the superiority of the proposed method, comparisons with other fault diagnosis methods in practical application are necessary. Our method consists of two main parts: the training paradigm (DINO) and the backbone of the encoder (ViT). Therefore, the contrast experiments also include different training paradigms and backbones. In terms of the training paradigm, conventional supervised learning and other self-supervised learning methods are involved. The supervised learning technique only considers the cross-entropy loss of the limited labeled data, and the self-supervised learning approaches adopted for comparisons incorporate the SimCLR \cite{pmlr-v119-chen20j} and BYOL \cite{NEURIPS2020_f3ada80d}. Then, in the aspect of the encoder backbone, ResNet18 \cite{7780459} and DenseNet121 \cite{8099726} are considered. 

These training paradigms and encoder backbones are used for the CWRU and XJTU datasets, respectively, and their results are summarized in Tab.\ref{tb:comparison}. Note that these fault diagnosis methods share the same data pre-processing pipeline shown in Fig.\ref{fig:preprocess}. Among them, due to the lack of labeled data, the self-supervised learning paradigm can not effectively train the networks. It is difficult to obtain the feature vectors with preferable generalization ability through self-supervised learning, so its testing accuracy is relatively low. By contrast, better fault diagnosis performance can be acquired by training the model network with the self-supervised learning method. Regardless of the encoder backbone, the testing accuracy obtained by DINO are generally higher than those of other self-supervised learning method, suggesting that DINO is a more powerful training paradigm. Besides, surprisingly, adopting the ResNet18 rather than ViT as the encoder backbone can get a higher testing accuracy on the XJTU dataset. But overall, among all these solutions, ViT trained by DINO achieves the highest average diagnosis accuracy on the given datasets, which further proves the effectiveness and superiority of the proposed method for the rolling bearing fault diagnosis with limited labeled data. 

\begin{table}[ht]
      \caption{Testing accuracy of the comparison methods on the CWRU and XJTU datasets}\label{tb:comparison}
      \begin{tabular*}{1.0\textwidth}{@{}LLLLL@{}}
            \toprule
            \multirow{2}*{\makecell[l]{Training \\ paradigm}} & \multirow{2}*{\makecell[l]{Encoder \\ backbone}} & \multicolumn{3}{c}{Accuracy} \\
            ~ & ~ & CWRU dataset & XJTU dataset & Average \\
            \midrule
            \multirow{3}*{Supervised} & ResNet18 & 77.41\% & 63.86\% & 70.64\% \\
            ~ & DenseNet121 & 72.34\% & 54.63\% & 63.49\% \\
            ~ & ViT & 46.86\% & 49.84\% & 48.35\% \\
            \midrule
            \multirow{3}*{SimCLR} & ResNet18 & 88.58\% & 82.75\% & 85.67\% \\
            ~ & DenseNet121 & 83.64\% & 87.92\% & 85.78\% \\
            ~ & ViT & 87.29\% & 67.16\% & 77.23\% \\
            \midrule
            \multirow{3}*{BYOL} & ResNet18 & 91.05\% & 83.09\% & 87.07\% \\
            ~ & DenseNet121 & 89.91\% & 83.94\% & 86.93\% \\
            ~ & ViT & 73.53\% & 63.84\% & 68.69\% \\
            \midrule
            \multirow{3}*{\textbf{DINO}} & ResNet18 & 93.56\% & \textbf{90.85\%} & 92.21\% \\
            ~ & DenseNet121 & 92.90\% & 85.83\% & 89.37\% \\
            ~ & \textbf{ViT} & \textbf{95.27\%} & 89.33\% & \textbf{92.30\%} \\
            \bottomrule
      \end{tabular*}
\end{table}

\section{Conclusions}

Faced with the contradiction between the conventional supervised fault diagnosis methods' dependency on massive labeled data and engineering practice, 
a Wavelet Transform (WT) and self-supervised learning-based fault diagnosis framework for bearing fault diagnosis with limited labeled data has been presented. In this method, the original vibration signals are pre-processed by WT and cubic spline interpolation to obtain the time-frequency maps (TFMs) with a specific scale. The teacher and student networks are established based on the Vision Transformer (ViT) encoder and projector head, and a pretext task called "local to global correspondence" is introduced for self-supervised learning. Adopting the Self-distillation with no labels (DINO) algorithm to train the teacher and student networks, fault-specific feature representation can be obtained. The main conclusions are summarized as follows. 

1) Through the DINO algorithm, effective feature vectors from the complex TFMs can be extracted by the trained ViT encoder, and involving the centering and sharpening operations in the teacher network can avoid the problem of mode collapse efficaciously during the self-supervised learning procedure. The complementary effect of the centering and sharpening is observed, where centering encourages over-uniformity but inhibits over-alignment, while sharpening has the opposite function. Only by applying both operations simultaneously can the mode collapse be avoided. 

2) In the situation that only 1\% labeled samples are included in the CWRU and XJTU datasets, adopting the feature vectors extracted by the trained encoder without any fine-tuning, 95.27\% and 89.33\% testing accuracy can be obtained based on the simple K-Nearest Neighbor (KNN) classifier.

3) The influence of hyperparameters is discussed in detail. Teacher temperature and student temperature can significantly affect the fault diagnosis performance. Inappropriate values of them may directly cause the mode collapse, so a careful adjustment is recommended. Then, the scale range and momentum parameter in centering are also crucial, and there is an optimal value for both hyperparameters. The computational complexity and diagnosis accuracy of the proposed framework increase with the expansion of the number of local crops, so it should be weighed according to the specific hardware environment in the actual application. Finally, the explicit law of the influence of the pseudo labels dimension is not observed, but on the whole, it has a relatively unapparent impact on the fault diagnosis accuracy of the proposed framework. 

4) Different training paradigms, such as supervised learning, SimCLR, BYOL, DINO, and various encoder backbones, including ResNet18, DenseNet121, and ViT, are considered in twelve comparative fault diagnosis approaches. Among them, the proposed method has the highest average accuracy on the CWRU and XJTU datasets, demonstrating its effectiveness and superiority. 

Since sufficient unlabeled data is needed in this study, the main limitation lies in that the proposed method is not suitable for the small sample. In future work, further research should focus on the improved diagnosis approach for small-sample learning. 

\section*{Acknowledgements}
It is very grateful for the financial supports from the National Major Science and Technology Projects of China (No. 2017-IV-0008-0045), the National Natural Science Foundation of China (Nos. 11972129, 11732005) and the Fundamental Research Funds for the Central Universities.











\printcredits

\bibliographystyle{model1-num-names}

\bibliography{cas-refs}



\end{document}